# Room-temperature photo-chromism of silicon vacancy centers in CVD diamond

Alexander Wood[1,2*], Artur Lozovoi[1*], Zi-Huai Zhang[3], Sachin Sharma[1], Gabriel I. López-Morales[1], Harishankar Jayakumar[1,4], Nathalie P. de Leon[3], and Carlos A. Meriles[1,5, †]

[1]Department. of Physics, CUNY-City College of New York, New York, NY 10031, USA. [2]University of Melbourne, Parkville VIC 3010, Australia. [3]Department of Electrical and Computer Engineering, Princeton University, Princeton, NJ 08544, USA. [4]University of Minnesota, Minneapolis, MN 55455, USA. [5]CUNY-Graduate Center, New York, NY 10016, USA.

**ABSTRACT**: The silicon-vacancy (SiV) center in diamond is typically found in three stable charge states, $SiV^0$, $SiV^-$ and $SiV^{2-}$, but studying the processes leading to their formation is challenging, especially at room temperature, due to their starkly different photo-luminescence rates. Here, we use confocal fluorescence microscopy to activate and probe charge interconversion between all three charge states under ambient conditions. In particular, we witness the formation of $SiV^0$ via the two-step capture of diffusing, photo-generated holes, a process we expose both through direct $SiV^0$ fluorescence measurements at low temperatures and confocal microscopy observations in the presence of externally applied electric fields. Further, we show that continuous red illumination induces the converse process, first transforming $SiV^0$ into $SiV^-$, then into $SiV^{2-}$. Our results shed light on the charge dynamics of SiV and promise opportunities for nanoscale sensing and quantum information processing.

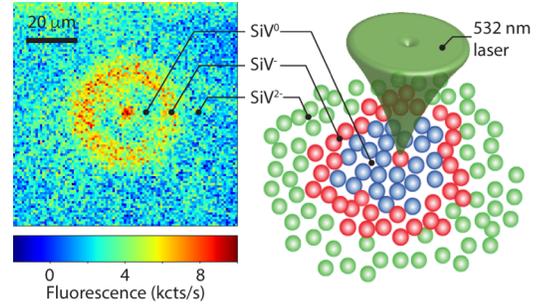

Spin-active color centers in wide bandgap semiconductors have emerged as a valuable platform for applications in quantum communication, computing, and sensing thanks to their long coherence lifetimes and optical access. Among the best-known examples are the nitrogen vacancy (NV) center in diamond and the di-vacancy center in silicon carbide, whose optical and spin properties have been the subject of extensive studies [1-5]. Part of this effort has also been devoted at identifying alternative color centers that can potentially outperform more established systems [6]. One illustration is the silicon vacancy (SiV) center in diamond, singled out in recent years as a promising spin qubit candidate with superior optical properties [7-9]. In particular, the neutral ($SiV^0$) and negatively-charged ($SiV^-$) states exhibit photo-chromism (with emission at 946 and 737 nm), possess optically-accessible electron spins ($S = 1$ and $1/2$), a high Debye-Waller factor, and an inversion symmetry that protects optical transitions from electric noise [10]. This allows their use close to the host crystal surface and reduces spectral diffusion of the zero-phonon line (ZPL), critical in quantum information processing and sensing applications. The SiV center can also be found in the doubly-negative state ($SiV^{2-}$), which is non-fluorescent and spin-less [11,12].

Among these alternative charge states, $SiV^0$ is of especial interest as it exhibits much longer spin coherence times than $SiV^-$ due to the absence of strong phonon-mediated spin relaxation processes [13,14]. First identified as the KUL1 center in chemical-vapor-deposition (CVD) diamond via electron paramagnetic resonance [15,16], $SiV^0$ was subsequently associated with a 946 nm (1.31 eV) zero-phonon line in the photoluminescence spectrum [17,18], only detectable below ~240 K in ensembles due to the suppression of otherwise dominant non-radiative processes. Long spin coherence and spin-lattice relaxation times have recently been observed below 20 K using pulsed electron paramagnetic resonance [14, 19] and optically detected magnetic resonance [20].

Despite much progress, stabilizing the SiV charge state into $SiV^0$ remains a challenging problem. First-principles calculations indicate that the preferred SiV charge states in synthetic diamond are $SiV^-$ or $SiV^{2-}$ [11,12]. It has been shown that boron doping during crystal growth shifts the Fermi level to favor the formation of $SiV^0$ [14]. Alternatively, hydrogen termination of the diamond surface has been exploited to induce the local formation of shallow $SiV^0$ centers with properties similar to those found in the bulk of the crystal [21].

Here we pay particular attention to processes of charge state conversion based on recombination of itinerant free carriers, which can be created via photoionization of nearby defects. This form of "photo-doping" — successfully demonstrated with NV centers [22,23] — has been shown to produce long-lived, non-equilibrium charge state distributions when the concentration of nitrogen impurities is moderate (~1 ppm or less) [24]. Subsequent experiments on SiVs showed that 532- or 632-nm illumination generates free carriers that are then captured outside the illumination region, resulting in the generation of a characteristic SiV$^-$ "bright halo" [25,26]. Identification of the charge carrier responsible for the observed effect is hindered by the fact that both SiV$^0$ and SiV$^{2-}$ are dark at room temperature. Experiments in the presence of external electric fields suggested that electron capture by SiV$^0$ plays a key role in the formation of the halo [26]. It was later argued, however, that annealed CVD diamond favors the formation of SiV$^{2-}$, not SiV$^0$, as the predominant charge state [27]. The notion of SiV$^{2-}$ as the charge state "by-default" gained support in recent experiments exploiting individual NV centers as charge sensors to reveal the sign of carriers photo-activated from proximal SiVs [28].

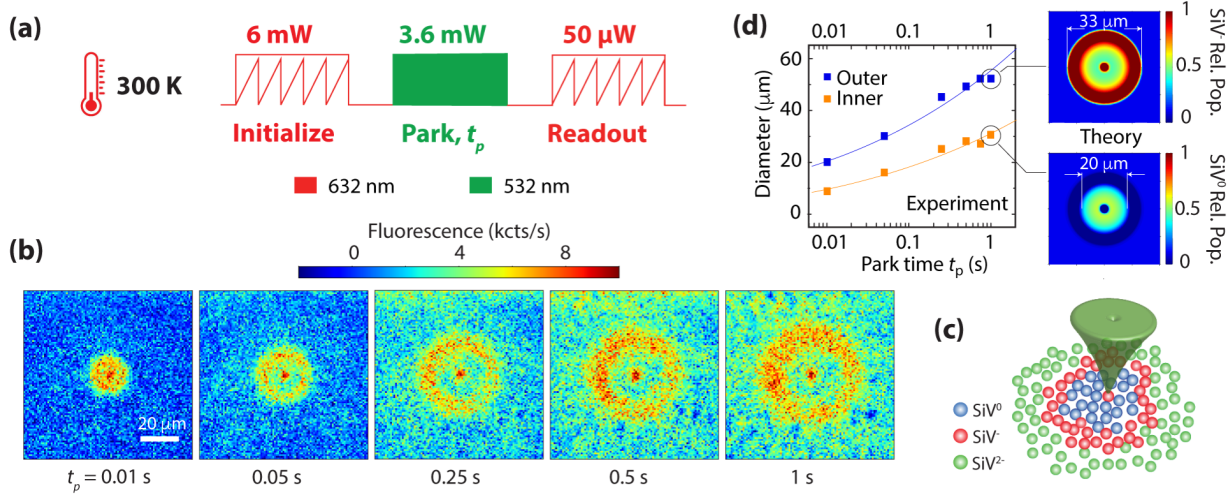

**Figure 1.** Photo-induced formation of SiV$^0$ under ambient conditions. (a) SiV$^-$ preparation and detection protocol at room temperature; upon a red laser scan (632 nm at 6 mW, 80×80 μm$^2$) and a green park (532 nm at 3.6 mW) for a variable time $t_p$, we image the resulting SiV$^-$ distribution via a weak red laser scan (632 nm at 50 μW, 10 ms per pixel, 100×100 pixels). (b) SiV$^-$ images as derived from the protocol in (a). (c) Schematics of charge dynamics under a green laser park; the inner blue ring signals the formation of SiV$^0$ via double hole capture. (d) (Left) Time dependence of the outer and inner radii in the measured charge patterns; solid lines are guides to the eye. (Right) Calculated SiV$^-$ and SiV$^0$ relative populations (top and bottom plots, respectively) for the conditions in (b) assuming a park time of 1 s.

In this work, we identify the conditions for the formation of all three SiV charge states — SiV$^0$, SiV$^-$ and SiV$^{2-}$ — under ambient conditions. We use multicolor confocal microscopy to image the SiV$^-$ charge state distribution at room temperature after exposure to photo-generated charge carriers, and confirm the inferred charge states by direct detection of SiV$^0$ photo-luminescence at 10 K. Further, we employ electric fields to probe the effect of drifting carriers, which allows us to discriminate between their signs and corresponding capture processes. Finally, we examine the response under continuous red illumination and show the sequential, laser-induced transformation of SiVs from neutral, to single-, to double-negatively charged. These room-temperature results are consistent with and complement observations under cryogenic conditions discussed in a recent study [29].

In our experiments, we use a CVD-grown [100] diamond with a nitrogen concentration of ~1 ppm as reported by the manufacturer (DDK); from prior studies [26], we estimate the SiV and NV concentrations at ~0.3 and ~0.03 ppm, respectively. For room-temperature measurements, we resort to a home-built confocal microscope accommodating 532- and 632-nm excitation laser paths; we use a bandpass filter to limit photon collection to a narrow window around the SiV$^-$ ZPL at 737 nm (see Supporting Information (SI), Section I).

The schematic in Fig. 1a lays out our experimental protocol: We first implement multiple 632-nm, 6-mW laser scan across a (80 μm)$^2$ area to initialize SiVs into a non-fluorescing charge state (SI, Section II). We subsequently park a 532-nm, 3.6 mW laser at the center of the scanned plane for a variable time $t_p$; illumination at this wavelength is known to generate a continuous stream of free electrons and holes stemming from charge cycling of co-existing NVs [30-32]. The ensuing carrier diffusion and capture produces a non-local change of the SiV charge state, which we subsequently image under 632-nm, 50-μW excitation.

Fig. 1b shows the results for variable park times: Consistent with prior observations [26], we first see a bright SiV$^-$ halo centered around the point of illumination after only 10 ms of green excitation. Upon increasing $t_p$, however, we witness the formation of a dark inner ring growing concomitantly with the outer bright pattern. This alternating, concentric structure suggests a double carrier capture process where SiVs not directly exposed to the green laser become negative (and hence bright) after a first capture event, to subsequently turn dark again upon capture of a second carrier. Provided the initial red scan preferentially produces SiV$^{2-}$, this process amounts to a double hole capture, and correspondingly implies the formation of an SiV$^0$-rich ring in the region adjacent to the parked beam (see schematics in Fig. 1c). Interestingly, preceding experiments missed these dynamics, instead reporting the formation of a uniformly bright SiV$^-$ disk [26]. We later show this is a consequence of the system's high sensitivity to the probe laser intensity, rapidly converting SiV$^0$ into SiV$^-$ even at these lower powers (see also SI, Section III).

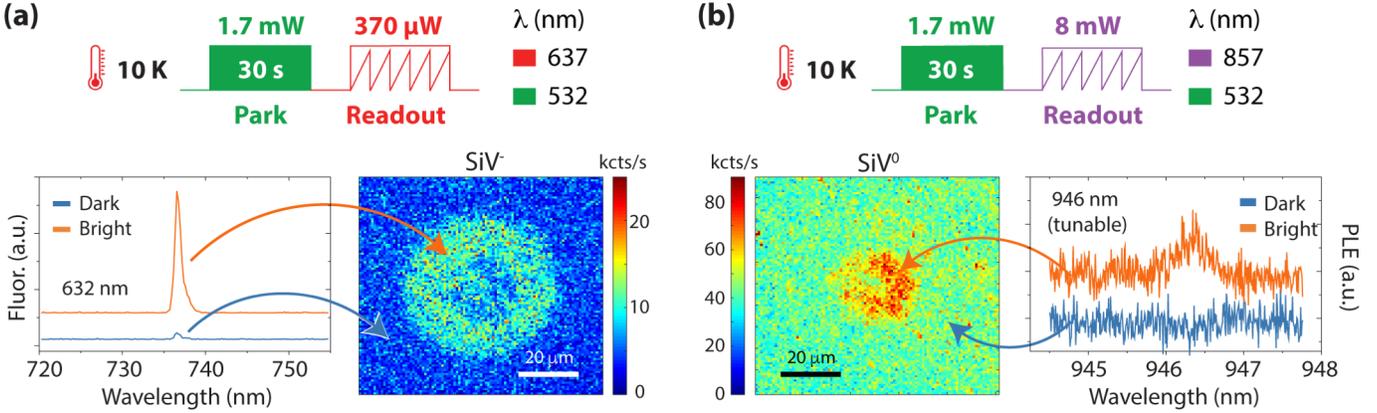

**Figure 2.** Direct observation of SiV$^0$ formation via hole capture. (a) (Top) SiV$^-$ imaging protocol at cryogenic temperatures. Following a 532-nm, 1.7-mW beam park at the center for 30 s, we image the resulting SiV$^-$ distribution using a 637-nm, 400-μW beam. (Bottom) Confocal imaging of SiV$^-$; the orange (blue) traces in the side photo-luminescence spectra shows that the bright (dark) areas of the image are SiV$^-$-rich (SiV$^-$-depleted). (b) Same as in (a) but for SiV$^0$. In this case, however, we use an 857-nm laser for image readout and obtain a photo-luminescence (PLE) spectrum via a 0.3 mW laser tunable around 946 nm. The bright ring at the image center directly demonstrates formation of SiV$^0$. In (a) and (b), spectra have been displaced vertically for clarity; all experiments at 10 K.

Conceptually, the formation of the observed SiV charge pattern must emerge from an interplay between carrier photo-activation, diffusion, and capture in the presence of localized optical excitation. This interplay can, in turn, be cast in terms of a set of master equations that simultaneously take into account the presence of SiV, NV, and substitutional nitrogen (N$_s$) in their different charge states (SI, Section III). Indeed, N$_s$ impurities — significantly more abundant than SiVs and NVs in CVD diamond [33] — charge-cycle between neutral and positively-charged when exposed to a stream of electrons and holes, hence playing an important role in the SiV pattern formation [34,35]. We capture these dynamics in Fig. 1d where we compare our observations — distilled in the main plot as the experimentally measured time evolution of the outer and inner ring diameters — with numerical simulations for a 1-s park time, shown in the right insert. Consistent with our model, we find the formation of an SiV$^0$-rich inner ring separated from the SiV$^{2-}$ background by an SiV$^-$ bright halo. A close inspection of the observed and calculated diameters, however, shows that the agreement is far from complete and must, therefore, be seen as qualitative. This is largely a consequence of the uncertainty surrounding the charge-state-dependent electron and hole capture cross sections for NVs, SiVs, and Ns, which in several cases are not known (SI, Section IV).

To disambiguate the composition of the inner ring, we carry out low temperature experiments where the SiV$^0$ fluorescence can be monitored directly [14]. To this end, we make use of a second cryo-workstation-based confocal microscope simultaneously integrating 532-, 637-, and 857-nm lasers (SI, Section I). With a wavelength above the SiV$^0$ recombination threshold at ~825 nm [20,36], 857-nm light facilitates non-destructive detection of SiV$^0$ fluorescence (too weak at room temperature due to non-radiative decay). Figs. 2a and 2b respectively display SiV$^-$- and SiV$^0$-selective images emerging upon a prolonged (60 s) green laser park time: We find complementary fluorescence maps that confirm the distinct charge state predominant in each section of the pattern. Further, optical spectroscopy on dark and bright sectors of either image directly indicate the formation of SiV$^0$ as revealed by the emission peak near 946 nm (upper trace in the side plot of Fig. 2b).

While the experiments above rely on itinerant carriers spreading from the point of photo-activation, the application of external electric fields allows us, in principle, to imprint a distinct spatial bias in the diffusion of electrons and holes, in principle observable at room temperature. We lay out this class of experiments in Fig. 3a, where we introduce a constant voltage difference across the imaged plane throughout the duration of the laser park. Fig. 3b compares our observations with and without external fields: We find that the SiV$^-$ ring previously observed under similar conditions (Fig. 1b) reshapes into a long tail pointing toward the negative electrode. Further, although the inner dark ring undergoes a less dramatic transformation, a close examination also indicates bias toward the upper section of the plane, all of which is consistent with the notion of hole trapping as the driving mechanism underlying SiV$^-$ formation. For completeness, we note that the observed changes in the outer sections of the fluorescence pattern — turned dark by the red initialization scan — are simply the (undesired) effect of scattered light on the red-initialized SiV ensemble and can, therefore, be ignored (SI, Section V).

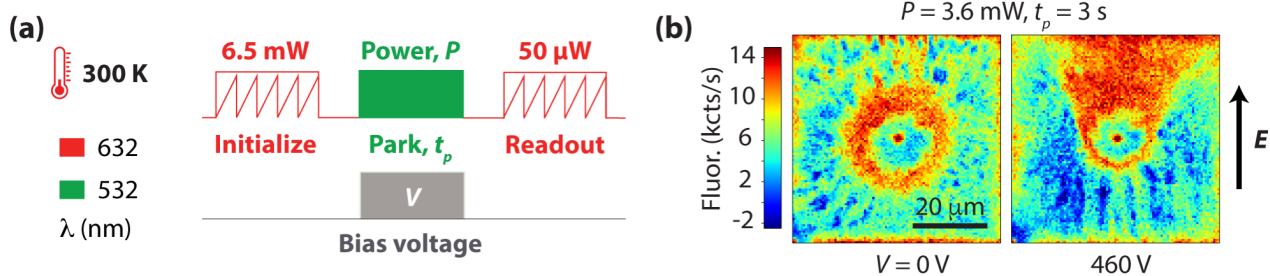

**Figure 3.** SiV charge state dynamics in the presence of a bias electric field. (a) Experimental protocol; we use silver planar electrodes separated by a 100-μm gap (not shown) to produce a time-independent electric field across the imaged area during a 532-nm laser park of variable power $P$ and duration $t_p$. (b) Confocal images of SiV$^-$ with and without a bias voltage $V$ for long park times (right and left images, respectively); the comet-tail-shaped pattern emerging in the presence of an external bias can be understood as the result of holes drifting under the external electric field $\mathbf{E}$. Throughout these experiments, the green laser power and park duration respectively are $P = 3.6$ mW and $t_p = 3$ s.

At face value, the results in Fig. 3 contradict some of the conclusions from our prior work[26] where similar experiments suggested SiV$^-$ generation from electron capture by SiV$^0$. Although dissimilar experimental conditions — including a different charge-state preparation history, a stronger readout beam, and the use of a vertical, not horizontal, electrode set — make an in-depth comparison difficult, we suspect that one regime or the other may be dominant depending on the chosen parameter set. In particular, preliminary observations — not presented here for brevity — show that local optical excitation and externally applied electric fields can combine to produce a rich, though complex, phenomenology; this includes, e.g., the formation of bi-directional, jet-like patterns where SiV$^-$ formation seemingly emerges both from hole capture by SiV$^{2-}$ in one half of the plane and from electron capture by SiV$^0$ in the other. We defer the in-depth discussion of these and related results — including an examination of the role of space charge fields in the SiV$^-$ pattern formation [37] — to an upcoming study.

Our ability to prepare SiV$^0$-rich ensembles at room temperature gives us the opportunity to probe their charge dynamics in the presence of continuous optical excitation [20,36]. To this end, we implement the experimental protocol in Fig. 4a: Upon creating an SiV$^0$-rich area, we park a 632-nm, 50-μW laser at select, off-centered positions for a variable time $t'_p$, to subsequently reconstruct an image using a weak red scan. Fig. 4b shows the result for a case where we sequentially illuminate eight different locations evenly distributed in the SiV$^0$ and SiV$^{2-}$ portions of the imaged plane (crosses in the upper left diagram). We find remarkably different responses: While red illumination in the outer sections of the pattern leads to sharp, dark spots, the opposite takes place near the center where we observe bright spots emerge. The latter signals a red-induced recombination of SiV$^0$ into SiV$^-$, whereas the former points to the recombination of SiV$^-$ (residually present in the SiV$^{2-}$ background after a green park, see SI, Section VI).

When combined, the observations above indicate a cascade process where red excitation first transforms SiV$^0$ into SiV$^-$, then into SiV$^{2-}$. We experimentally capture this full sequence by monitoring the fluorescence from a near-center section of the pattern upon a variable red park time (Fig. 4c and side image series in Fig. 4b). Following an early growth of SiV$^-$ fluorescence (0–1 s), we subsequently find a gradual decay that we interpret as a recombination into SiV$^{2-}$. This light-induced charge-state progression mirrors the two-step, diffusion-driven hole capture process highlighted before, and hence can be effectively seen as its converse.

Remarkably, prolonged red excitation on the SiV$^0$-rich background also results in the formation of a surrounding bright halo slowly growing over time to reach an outer diameter of ~15 μm after a 40-s park (image series in Fig. 4b). Given the starting charge state composition of the ensemble, we must attribute SiV$^-$ formation to electron capture at SiV$^0$ sites, presumably facilitated here by a stream of itinerant electrons produced by local $N_s^0$ (and NV$^-$) ionization under the 632-nm laser beam. Note that SiV$^0$ formation during the (preceding) green park is likely accompanied by the preferential transformation of $N_s^+$ (arguably abundant after the red initialization scan) into neutral. In the absence of competing trapping sites, one would correspondingly expect the probability of electron capture by SiV$^0$ to grow, consistent with our observations (Fig. 4d). This physical picture, of course, hinges on the assumption of a sufficiently large electron reservoir at the point of red excitation; additional, more-quantitative work will therefore be needed to gain an accurate description of the charge dynamics at play.

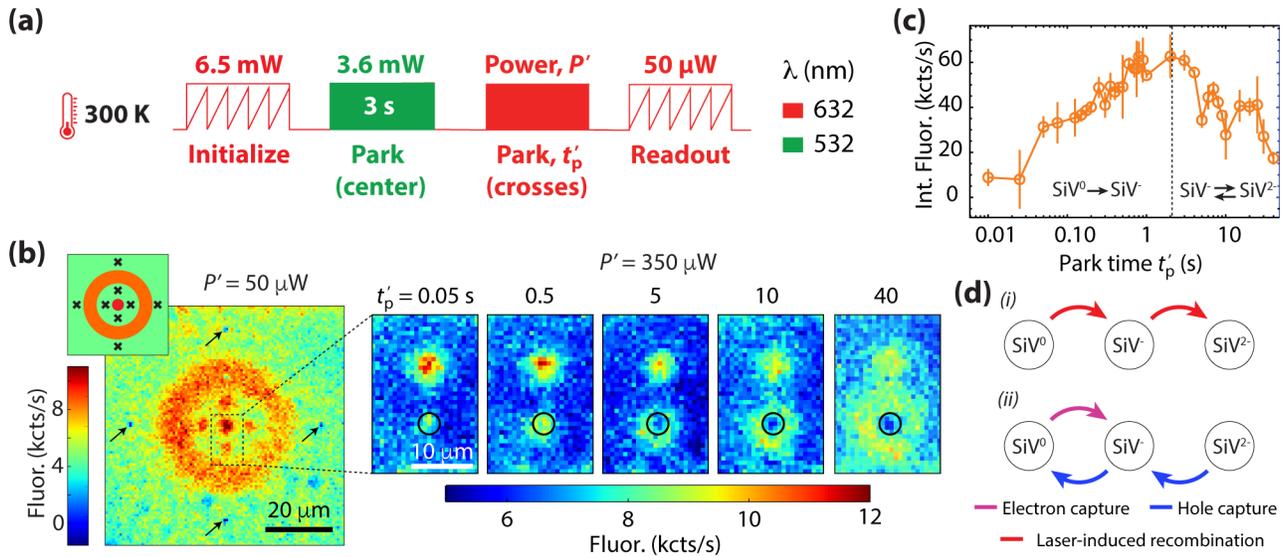

**Figure 4.** Charge dynamics of SiV under red illumination. (a) Experimental protocol. Following a 3-s green beam park on a red-laser-initialized background, we use a 632-nm laser to illuminate select points in the pattern. (b) Confocal image of SiV$^-$ (100×100 pixels, 10 ms per pixel) upon application of the protocol in (a). Crosses in the upper left insert indicate the locations of 632-nm, 50-μW parks relative to the bright ring in the pattern for a time $t'_p = 0.5$ s. Arrows indicate the positions of the outer parks, less visible than those in the inner dark ring. The image series on the right-hand side shows zoomed confocal maps of the area within the dashed square. We monitor the integrated SiV$^-$ fluorescence within the black circle for variable red laser park durations ($P' = 350$ μW). (c) Integrated SiV$^-$ fluorescence versus $t'_p$ at the inner (dark) ring position highlighted in (b); red excitation first creates, then partially depletes SiV$^-$. (d) Putative SiV charge dynamics under 632-nm illumination. (*i*) Upon direct exposure to red light, SiV$^0$ first recombines into SiV$^-$, then into SiV$^{2-}$. (*ii*) In the dark, SiV$^-$ formation arises from hole capture by SiV$^{2-}$ or from electron capture by SiV$^0$; one scenario or the other become dominant depending on the charge state initialization history.

In summary, our experiments in CVD diamond show that optical illumination combined with carrier diffusion and capture lead to the formation of fluorescence patterns segmented into areas where SiVs take different predominant charge states. In particular, we demonstrate the room-temperature formation of SiV$^0$-rich disks surrounding the point of local green illumination, which we associate to a sequential hole capture process, first transforming SiV$^{2-}$ into single-negatively charged, then into neutral. Starting with an SiV$^0$ ensemble, we show that the exact converse process can take place locally under continuous 632-nm illumination. Further, these same observations also suggest the alternative formation of SiV$^-$ from itinerant electron, not hole, capture, which manifests or not depending on the charge initialization history.

Our results open up opportunities for the overall characterization of SiV charge state dynamics (including, e.g., the identification of ionization/recombination thresholds) through the combined use of microscopy and optical spectroscopy techniques. The efficient hole-capture-assisted formation of SiV$^0$ we observe raises the question as to whether other hitherto hidden features can be similarly exposed at room-temperature. Of particular interest are processes such as optical spin polarization and/or spin-selective recombination, which could perhaps be exploited for optically detecting SiV$^0$ magnetic resonance under ambient conditions via spin-to-charge conversion protocols.

## ASSOCIATED CONTENT

**Supporting Information:** Contains additional information on sample characteristics and instrumentation used. It also provides additional details on the Monte Carlo modelling.

## AUTHOR INFORMATION

**Corresponding author:**
†E-mail: cmeriles@ccny.cuny.edu
Notes
(*) denotes equally contributing authors.
The authors declare no competing financial interests.

## ACKNOWLEDGMENTS

All authors acknowledge support by the U.S. Department of Energy, Office of Science, National Quantum Information Science Research Centers, Co-design Center for Quantum Advantage (C2QA) under contract number DE-SC0012704. A.W., A.L., S.S., H.J., and C.A.M. acknowledge access to the facilities and research infrastructure of NSF CREST-IDEALS, grant number NSF-HRD-1547830. A.W. was additionally supported by the Australian Research Council (Grant ID DE210101093).
**REFERENCES**

[1] Doherty, M.W., Manson, N.B., Delaney, P., Jelezko, F., Wrachtrup, J. and Hollenberg, L.C., The nitrogen-vacancy colour centre in diamond. *Phys. Rep.* **2013**, *528*, 1-45.

[2] Pompili, M., Hermans, S.L.N., Baier, S., Beukers, H.K.C., Humphreys, P.C., Schouten, R.N., Vermeulen, R.F.L., Tiggelman, M.J., dos Santos Martins, L., Dirkse, B., Wehner, S., and Hanson, R. Realization of a multinode quantum network of remote solid-state qubits. *Science* **2021**, *372*, 259-264.

[3] Bradley, C.E., Randall, J., Abobeih, M.H., Berrevoets, R.C., Degen, M.J., Bakker, M.A., Markham, M., Twitchen, D.J. and Taminiau, T.H., A ten-qubit solid-state spin register with quantum memory up to one minute. *Phys. Rev. X* **2019**, *9*, 031045.

[4] Christle, D.J., Klimov, P.V., Charles, F., Szász, K., Ivády, V., Jokubavicius, V., Hassan, J.U., Syväjärvi, M., Koehl, W.F., Ohshima, T., Son, N.T., Janzen, E., Gali, A. and Awschalom, D.D. Isolated spin qubits in SiC with a high-fidelity infrared spin-to-photon interface. *Phys. Rev. X* **2017**, *7*, 021046.

[5] Anderson, C.P., Glen, E.O., Zeledon, C., Bourassa, A., Jin, Y., Zhu, Y., Vorwerk, C., Crook, A.L., Abe, H., Ul-Hassan, J., Ohshima, T., Son, N.T., Galli, G. and Awschalom, D.D. Five-second coherence of a single spin with single-shot readout in silicon carbide. *Sci. Adv.* **2022**, *8*, eabm5912.

[6] Wolfowicz, G., Heremans, F.J., Anderson, C.P., Kanai, S., Seo, H., Gali, A., Galli, G. and Awschalom, D.D., Quantum guidelines for solid-state spin defects. *Nat. Rev. Mater.* **2021**, *6*, 906.

[7] Hepp, C., Müller, T., Waselowski, V., Becker, J.N., Pingault, B., Sternschulte, H., Steinmüller-Nethl, D., Gali, A., Maze, J.R., Atatüre, M. and Becher, C., Electronic structure of the silicon vacancy color center in diamond. *Phys. Rev. Lett.* **2014**, *112*, 036405.

[8] Sipahigil, A., Jahnke, K.D., Rogers, L.J., Teraji, T., Isoya, J., Zibrov, A.S., Jelezko, F. and Lukin, M.D. Indistinguishable photons from separated silicon-vacancy centers in diamond. *Phys. Rev. Lett.* **2014**, *113*, 113602.

[9] Sukachev, D.D., Sipahigil, A., Nguyen, C.T., Bhaskar, M.K., Evans, R.E., Jelezko, F. and Lukin, M.D. Silicon-vacancy spin qubit in diamond: a quantum memory exceeding 10 ms with single-shot state readout. *Phys. Rev. Lett.* **2017**, *119*, 223602.

[10] Dietrich, A., Jahnke, K. D., Binder, J. M., Teraji, T., Isoya, J., Rogers, L. J. and Jelezko, F. Isotopically varying spectral features of silicon-vacancy in diamond. *New J. Phys.* **2014**, *16*, 113019.

[11] Gali, A. and Maze, J. R. Ab initio study of the split silicon-vacancy defect in diamond: Electronic structure and related properties. *Phys. Rev. B* **2013**, *88*, 235205.

[12] Thiering, G. and Gali, A. Ab Initio Magneto-Optical Spectrum of Group-IV Vacancy Color Centers in Diamond. *Phys. Rev. X* **2018**, *8*, 021063.

[13] Jahnke, K.D., Sipahigil, A., Binder, J.M., Doherty, M.W., Metsch, M., Rogers, L.J., Manson, N.B., Lukin, M.D. and Jelezko, F. Electron–phonon processes of the silicon-vacancy centre in diamond. *New J. Phys.* **2015**, *17*, 043011.

[14] Rose, B.C., Huang, D., Zhang, Z.H., Stevenson, P., Tyryshkin, A.M., Sangtawesin, S., Srinivasan, S., Loudin, L., Markham, M.L., Edmonds, A.M., Twitchen, D.J., Lyon, S.A. and De Leon, N.P. Observation of an environmentally insensitive solid-state spin defect in diamond. *Science* **2018**, *361*, 60.

[15] Iakoubovskii, K., Stesmans, A., Nouwen, B. and Adriaenssens, G. J. ESR and optical evidence for a Ni vacancy center in CVD diamond. *Phys. Rev. B* **2000**, *62*, 16587.

[16] Edmonds, A. M., Newton M. E., Martineau P. M., Twitchen D. J. and Williams S. D. Electron paramagnetic resonance studies of silicon-related defects in diamond. *Phys. Rev. B* **2008**, *77*, 245205.

[17] D'Haenens-Johansson, U. F. S., Edmonds, A. M., Newton, M. E., Goss, J. P., Briddon, P. R., Baker, J. M., Martineau, P. M., Khan, R. U. A., Twitchen, D. J. and Williams, S. D. EPR of a defect in CVD diamond involving both silicon and hydrogen that shows preferential alignment. *Phys. Rev. B* **2010**, *82*, 155205.

[18] D'Haenens-Johansson, U.F.S., Edmonds, A.M., Green, B.L., Newton, M.E., Davies, G., Martineau, P.M., Khan, R.U.A. and Twitchen, D.J. Optical properties of the neutral silicon split-vacancy center in diamond. *Phys. Rev. B* **2011**, *84*, 245208.

[19] Green, B.L., Mottishaw, S., Breeze, B.G., Edmonds, A.M., D'Haenens-Johansson, U.F.S., Doherty, M.W., Williams, S.D., Twitchen, D.J. and Newton, M.E. Neutral silicon-vacancy center in diamond: Spin polarization and lifetimes. *Phys. Rev. Lett.* **2017**, *119*, 096402.

[20] Zhang, Z.H., Stevenson, P., Thiering, G., Rose, B.C., Huang, D., Edmonds, A.M., Markham, M.L., Lyon, S.A., Gali, A. and De Leon, N.P. Optically detected magnetic resonance in neutral silicon vacancy centers in diamond via bound exciton states. *Phys. Rev. Lett.* **2020**, *125*, 237402.

[21] Zhang, Z.H., Zuber, J.A., Rodgers, L.V., Gui, X., Stevenson, P., Li, M., Batzer, M., Grimau, M., Shields, B., Edmonds, A.M., Palmer, N., Markham, M.L., Cava, R.J., Maletinsky, P. and De Leon, N.P. Neutral silicon vacancy centers in undoped diamond via surface control. *arXiv:2206.13698* (2022).

[22] Jayakumar, H., Henshaw, J., Dhomkar, S., Pagliero, D., Laraoui, A., Manson, N.B., Albu, R., Doherty, M.W. and Meriles, C.A. Optical patterning of trapped charge in nitrogen-doped diamond. *Nat. Commun.* **2016**, *7*, 1-8.

[23] Dhomkar, S., Henshaw, J., Jayakumar, H. and Meriles, C.A., Long-term data storage in diamond. *Sci. Adv.* **2016**, *2*, e1600911.

[24] Collins, A.T. The Fermi level in diamond. *J. Phys. Cond. Matter* **2002**, *14*, 3743.
6

# Supporting Information for

## "Room-temperature photo-chromism of silicon vacancy centers in CVD diamond"

Alexander Wood[1,2*], Artur Lozovoi[1*], Zi-Huai Zhang[3], Sachin Sharma[1], Gabriel I. López-Morales[1], Harishankar Jayakumar[1,4], Nathalie P. de Leon[3], and Carlos A. Meriles[1,5,†]

[1]Department. of Physics, CUNY-City College of New York, New York, NY 10031, USA. [2]University of Melbourne, Parkville VIC 3010, Australia. [3]Department of Electrical and Computer Engineering, Princeton University, Princeton, NJ 08544, USA. [4]University of Minnesota, Minneapolis, MN 55455, USA. [5]CUNY-Graduate Center, New York, NY 10016, USA.

## I. Experimental apparatus

Room temperature and cryogenic measurements are carried out with the help of two different fluorescence confocal microscopes:

### I.1 *Room-temperature microscope*

This home-built scanning system features green and red excitation lasers and removable bandpass filters to select regions of interest in the emission of NV and SiV centers in diamond. The complete system is depicted in Fig. S1: Typical operating parameters and key components include a 40-mW laser at 532 nm sourced from a DPSS laser diode module (Thorlabs, DJ532-40) and 633-nm red light from a 70-mW laser diode (Thorlabs, HL63163DG), each driven by home-made laser diode drivers. A mechanical shutter gates the 532-nm beam while a TTL signal into the driver board gates the red laser current. Light from both sources is combined via a dichroic mirror and coupled into a single-mode fiber (Thorlabs P3-460B-FC-2). The output from the fiber is collimated with an NA = 0.5 achromatic objective lens (Olympus UMPlanFl N). The scanning confocal microscope uses a two-axis galvo steering mirror system (Thorlabs GVS002), a 4$f$-relay lens ($f_1 = 100$ mm, $f_2 = 200$ mm), and a NA = 0.4 microscope objective (Mitutoyo MPlan Apo 2) mounted on a one-axis piezo scanning stage (PI P-725.2CA).

Light from the diamond is collected using the reverse optical path and subsequently separated from the excitation light by a dichroic mirror (Thorlabs DMLP-638). A series of filters is used to select regions of the PL spectrum: 700 nm and 650 nm long-pass filters prevent red light leakage to the detection, while a 735-nm bandpass filter selects out the SiV$^-$ ZPL. The PL is then focused using a 12.5-mm aspheric lens into a single-mode fiber (MFD = 10 µm) and directed into a single-photon counting module (SPCM, Excelitas SPCM-AQRH-14). Photon counts are recorded using a National Instruments NI-PCIe 6321 data acquisition card, which additionally supplies the analog voltages for driving the galvo scanning system.

To apply electric fields, we deposit a thin layer (<100 nm) of silver metal onto the diamond surface in the shape of two rectangular pads separated by 100 µm. These electrodes are connected to the output terminals of a manually-controlled bank of DC power supplies, capable of generating a voltage of +/-460 V across the electrodes.

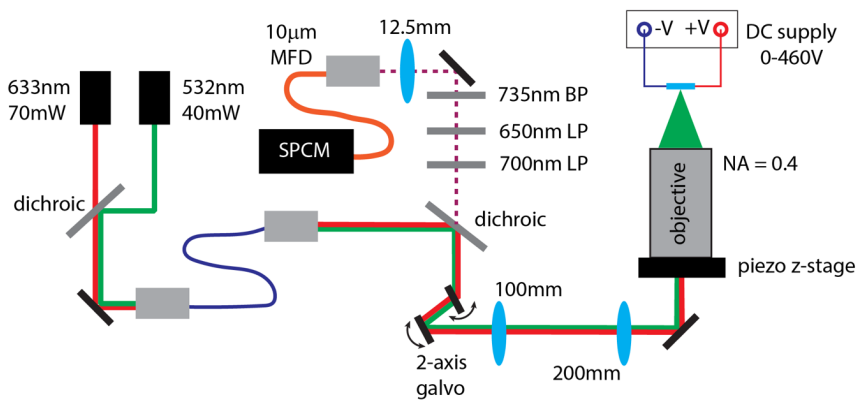

**Figure S1.** Room-temperature setup at CCNY, showing key components and optical paths.

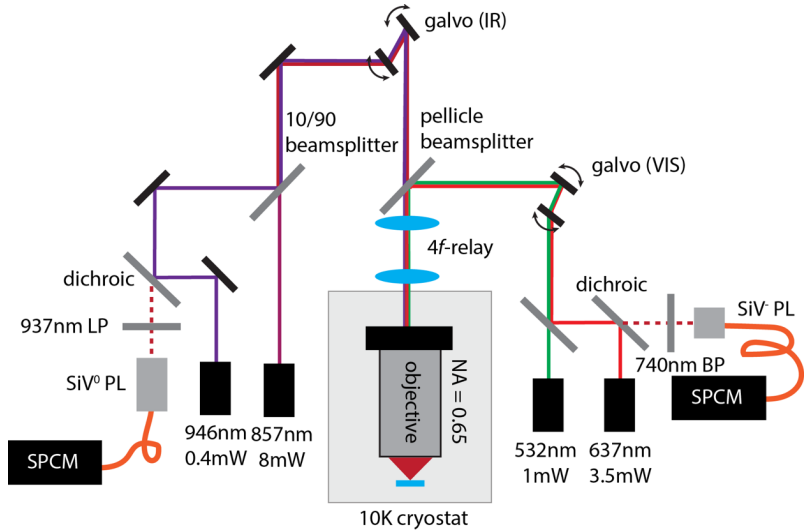

**Figure S2.** Experimental dual-galvo setup at Princeton University, showing key components and optical paths.

**I.2** *Low-temperature microscope*

We conduct fluorescence measurements at 10 K via a home-built dual-galvo scanning confocal microscope with two independent branches for detection and excitation of $SiV^-$ and $SiV^0$. The complete system is schematically shown in Fig. S2. The sample sits within a helium flow cryostat (Janis ST-500 probe station). Both branches contain a scanning galvo system (Thorlabs GVS012) and are combined using a pellicle beamsplitter (Thorlabs BP245B1). After that, the beam is sent through a $4f$ system into a NIR 50× objective lens (Olympus LCPLN50XIR) that is placed inside vacuum within the cryostat.

In the $SiV^0$ branch, excitation and detection channels are combined using a 925-nm dichroic beam-splitter (Semrock FF925-Di01-25-D). Resonant illumination for photoluminescence excitation at the ZPL is provided by a tunable diode laser (Toptica CTL 950) and is combined with the detection channel using a 10/90 beamsplitter (Thorlabs BS044). Photoluminescence is detected in the emission sideband with a 980-nm long-pass filter (Semrock LP02-980RE-25). Off-resonant excitation is performed with a tunable diode laser (Toptica DL pro 850 nm) and the photoluminescence is filtered using a 937-nm long pass filter (Semrock FF01-937/LP-25). The laser is pulsed using a home-built shutter system, and intensity-controlled by a variable optical attenuator (Thorlabs V800PA). Finally, the photoluminescence is coupled into a single mode fiber and detected either by a CCD spectrometer (Princeton Instruments Acton SP-2300i with Pixis 100 CCD and 300 g/mm grating) or by a superconducting nanowire detector (Quantum Opus, optimized for 950 - 1100 nm).

In the $SiV^-$ branch, excitation and detection channels are combined with a 650-nm dichroic beam-splitter (Thorlabs DMLP650). The excitation channel is equipped with 532-nm (Lambda Pro UG-100 mW) and 637-nm (Thorlabs LP637-SF70) lasers that are coupled to a single fiber using a RGB combiner (Thorlabs RGB26HF). The 637-nm laser is pulsed using a home-built shutter. The 532-nm laser is pulsed using an acousto-optic modulator and the intensity is controlled by a variable optical attenuator (Thorlabs V600A). Photoluminescence is filtered using a 740-nm bandpass filter (Thorlabs FB740-10) and is detected by a single photon detector (Excelitas SPCM-AQRH-44-FC).

**II. Initialization protocols**

We initialize the NV and SiV charge states prior to parking the laser beam via a laser scan. In general, the processes at work in these scans can be complex, requiring careful selection of laser power, scanning speed and number of repeats to optimize the initialization fidelity. Some components of the dynamics at play can be singled out, such as the photo-ionization of NV centers to the dark $NV^0$ state, which is accomplished with good fidelity using only a single high-power red scan. Preparation of SiV in a well-defined charge state is more involved, however, since the scanning laser beam continuously generates charges that are captured by remote defects, while simultaneously undergoing direct photo-excitation under the illumination spot. For instance, $SiV^{2-}$ is rapidly generated under direct red illumination, while $SiV^-$ is formed from proximal capture of holes in the region immediately adjacent to the point of illumination. Consequently, $SiV^{2-}$ initialization is optimized when a high power (6.5 mW) red laser is rapidly scanned (i.e., a short dwell time per pixel) multiple times. Figures S3a and S3b show some characteristic $SiV^-$ response as a function of the number of red scans and red laser power.

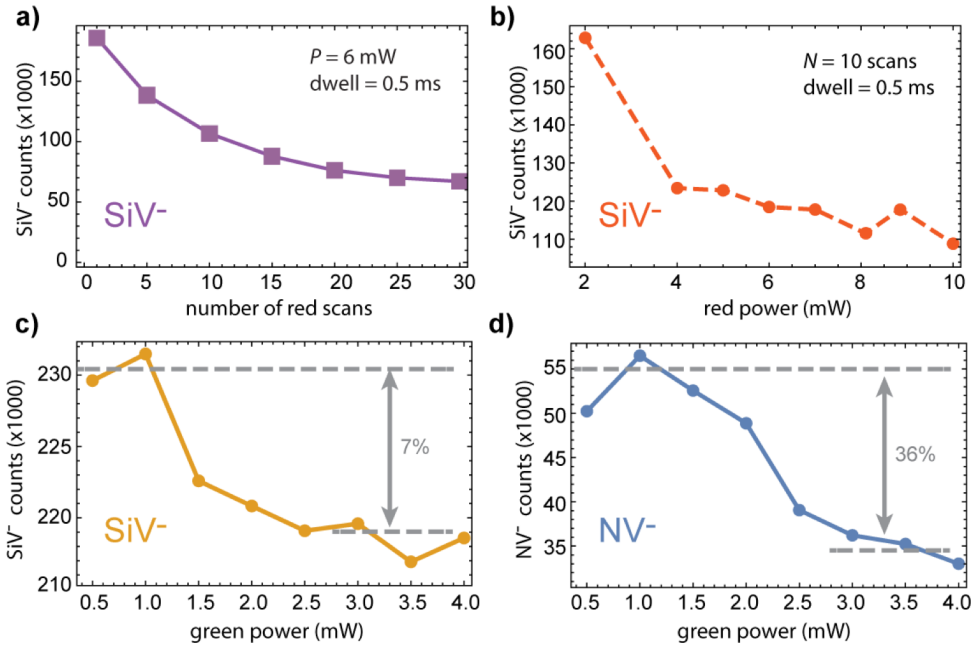

**Figure S3.** Optimizing red and green initialization scans at room temperature. (a) For a fixed red power (6.5 mW) and dwell-time-per-pixel (0.5ms), background SiV⁻ fluorescence (mean photon counts in a given image area) are reduced as the scan number increases; up to 30 scans are required to bleach a region of interest into SiV$^{2-}$. (b) Power dependence of red bleaching scans for 10 scans and 0.5-ms dwell time. (c) SiV⁻ and (d) NV⁻-band fluorescence as a function of green power after 1 scan with a 1-ms dwell time. Note the significant reduction of NV⁻ fluorescence, indicating population of NV$^0$: a significant amount of NV fluorescence is also collected in the SiV⁻ band, and competes with an increase in SiV⁻ at higher powers, resulting in a smaller (7%) drop in mean counts.

Green laser power offers a rudimentary though effective handle on selecting which charge states are preferentially initialized in a green scan. In Figs. S3c and S3d we plot the SiV⁻ and NV⁻ fluorescence as a function of green laser power, showing that under higher green laser powers, SiV⁻ remains abundant, in agreement with Ref. [1]. The slight (7%) reduction of SiV⁻ fluorescence in Fig. S3c is mostly an artifact due to a reduction of the NV fluorescence captured within the SiV⁻ band, which typically exceeds the SiV⁻ ZPL emission. Table 1 concisely summarizes the optimal parameters required for generating a particular charge state using green or red laser scanning.

Under our present conditions, the generation of SiV$^0$ is not strongly affected by the charge state initialization history. Figures S4a and S4b show the SiV⁻ pattern following a green park on SiV⁻- and SiV$^{2-}$-rich surfaces produced via green and red laser scans, respectively. In both cases, we find dark central disks similar in size, which indicates the formation of SiV$^0$ remains largely unchanged. Given the greater abundance of substitutional nitrogen — and, thus, its dominant role on the charge dynamics of the system — this observation suggests that green and (repeated) red illumination have a comparable effect on N$_s$, preferentially transforming into positively charged. For completeness, we mention that similar observations in the NV⁻-selective band show that the dark NV$^0$ disk matches the outer diameter of the SiV⁻ ring in Fig. S4b, as would be expected for a single hole capture process.

| Defect/charge state | Laser λ (nm) | Power (mW) | Dwell (ms) | N scans |
|---|---|---|---|---|
| NV⁻ | 532 | 0.3 | 1 | 1 |
| NV$^0$ | 633 | 6.5 | 0.5 | 1 |
| SiV⁻ | 532 | 3.6 | 1 | 1 |
| SiV$^{2-}$ | 633 | 6.5 | 0.5 | 30 |

**Table S1.** Parameters used in scans to selectively initialize region to a given defect charge state.

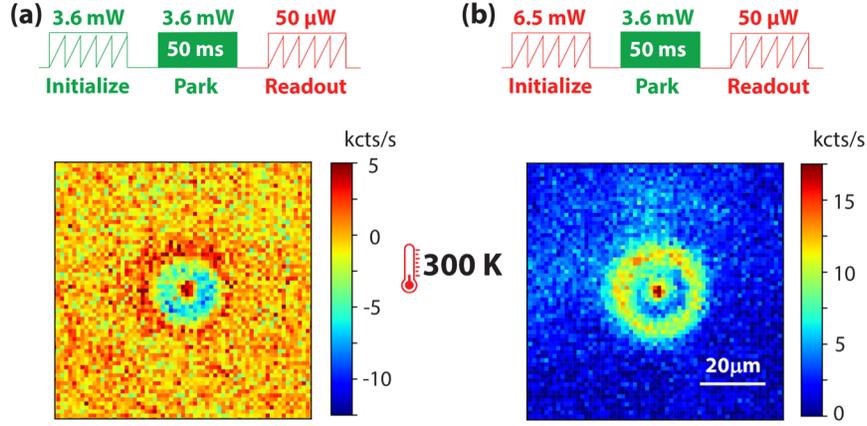

**Figure S4.** (a) Experimental protocol used to generate a dark SiV$^0$ halo after a 532-nm, 3.6-mW park during 50 ms on a bright, SiV$^-$-rich, background, which is initialized using a 532-nm, 3.6-mW scan (1 ms dwell time). (b) Same as (a), but for a dark background, which is initialized by repeating a 632-nm, 6.5-mW scan (0.5 ms dwell time) for $N = 30$ times. Under these conditions, we regain the SiV$^0$ / SiV$^-$ concentric structure discussed in the main text.

### III. Dependence on green and red laser powers

Figure 1b of the main text depicts the time dependence of the SiV$^-$ pattern. We have also examined the dependence under a fixed time and variable green laser power. Figure S5 depicts the relevant data, taken with a 1-s green laser park on a background red-initialized to the SiV$^{2-}$ state.

Readout of the SiV$^0$ charge state via photo-recombination is also highly power dependent and is potentially a reason why its observation at room temperature has remained elusive (for example, being absent in Ref. [4]). Figure S6 shows readout images after identical green initializations and green laser parks, but differing 633-nm readout powers. With a readout power of 50 µW, the dark inner disk is clearly resolved, but at higher powers (400 µW) a single bright halo appears instead: In this particular experiment, the 400-µW red light cumulatively bleaches during the reference and imaging scans, resulting in a significantly darker background for the higher readout power.

### IV. Numerical simulations

To describe the effect of diffusing photo-generated carriers on the SiV charge state, we solve a set of master equations capturing the time evolution in the concentration of holes ($p(r,t)$), electrons ($n(r,t)$), doubly negatively charged SiV$^{2-}$ ($S_{2-}(r,t)$), negatively charged SiV$^-$ ($S_-(r,t)$), negatively charged NV$^-$ ($Q_-(r,t)$) positively charged N$_s^+$ ($P_+(r,t)$) in the presence of optical excitation. The total volume concentration of SiV, NV and N$_s$ are $S$, $Q$, and $P$, respectively. We consider the processes of hole and electron diffusion, hole capture by SiV⁻, SiV²⁻, NV⁻, and $N_s^0$, electron capture by SiV⁰ and $N_s^+$, photoionization of $NV^-$ and $N_s^0$, and photo-recombination of $NV^0$, $SiV^0$, and SiV⁻. The complete set of equations is given by:

$$\frac{\partial Q_-}{\partial t} = (Q - Q_-)\vartheta_0 - Q_- - \sigma_{NVp} p Q_- ,$$

$$\frac{\partial P_+}{\partial t} = (P - P_+)\vartheta_N - \sigma_{Nn} n P_+ + \sigma_{Np} p (P - P_+),$$

$$\frac{\partial S_{2-}}{\partial t} = S_- \vartheta_{Si-} - \sigma_{pS2-} p S_{2-} ,$$

$$\frac{\partial S_-}{\partial t} = S_- \vartheta_{Si-} + \sigma_{pS2-} p S_{2-} - \sigma_{pS-} p S_- + \sigma_n n (S - S_- - S_{2-}) + \vartheta_{Si0}(S - S_- - S_{2-}) ,$$

$$\frac{\partial n}{\partial t} = D_n \nabla^2 n + Q_- \vartheta_- + (P - P_+)\vartheta_N - \sigma_{Nn} n P_+ - \sigma_{nS0} n (S - S_- - S_{2-}),$$

$$\frac{\partial p}{\partial t} = D_p \nabla^2 p + (Q - Q_-)\vartheta_0 - \sigma_{NVp} p Q_- + \sigma_{Np} n P_+ - \gamma_p p(P - P_+) + S_- \vartheta_{Si-} - \sigma_{pS-} p S_- - \sigma_{pS2-} p S_{2-} + \vartheta_{S0}(S - S_- - S_{2-})$$

where $\nabla^2$ denotes the Laplace operator in two dimensions. Photo-recombination and ionization under 532 nm illumination are modelled assuming the beam has a diffraction-limited Gaussian profile with the ionization rates expressed as $\vartheta \cdot \left(\frac{I}{I_0}\right)^2 \cdot \exp\left(-\frac{r^2}{s^2}\right)$ for NV and $\vartheta \cdot \frac{I}{I_0} \cdot \exp\left(-\frac{r^2}{s^2}\right)$ for $N_s$ in agreement with a two- and one-photon process, where $I$ = 3.6 mW, $I_0$ = 1 µW, $s$ = 1 µm. SiV recombination rates are taken to be one-photon processes with $\vartheta_{S0}$ being similar to NV⁻ photoionization and $\vartheta_{S-}$ - one order of magnitude lower, in reasonable agreement with the experimental results shown in the main text. The SiV⁻ and SiV²⁻ hole capture cross sections, $\sigma_{pS-}$ and $\sigma_{pS2-}$ are taken to be 0.02× and 20× that of the NV⁻ hole capture cross section, $\sigma_{NVp}$. The SiV⁰ electron capture cross section, $\sigma_{nS0}$, is taken to be the same as the $N_s^0$ hole capture cross section, $\sigma_{Nn}$. Under these conditions, the correspondence in relative diameters of the inner and outside rings between the simulation and the experiments was found to be the best. The whole set of values used in the calculations is summarized in Table S2 along with literature references. We solve this system of equations in 1D using the Matlab parabolic differential equation solver, and then project the results onto the imaged plane taking advantage of the radial symmetry of the problem.

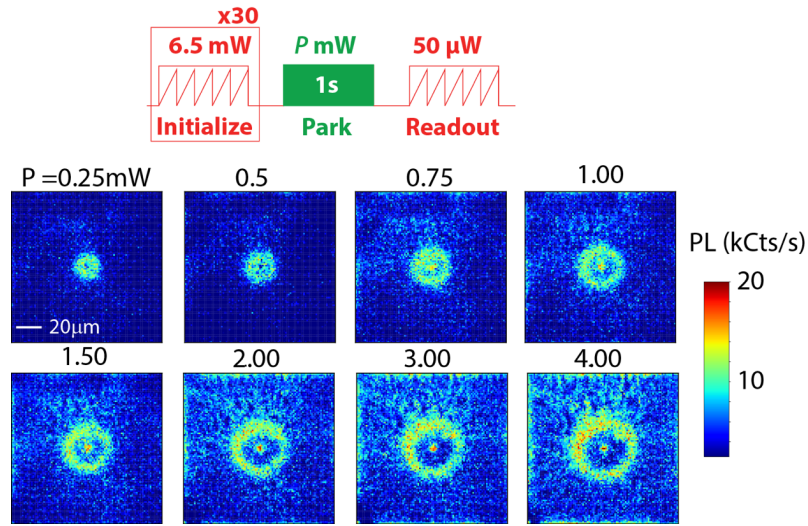

**Figure S5.** Power dependence of SiV$^0$/SiV$^-$ double disk structure. The upper panel depicts the background initialization and halo generation protocol. A fixed duration, 1-s green park is used, and the power varied from 0.25 mW to 4 mW. The double-halo feature is observed to grow with laser power, a characteristic feature of charge carrier generation and capture.

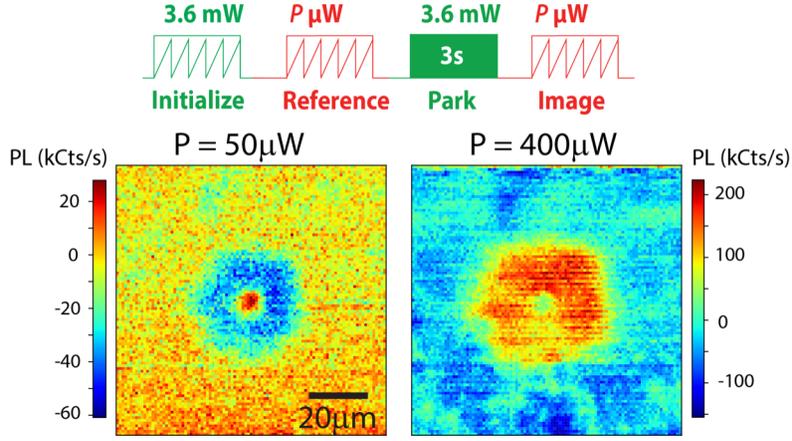

**Figure S6.** Red readout power dependence. Upper panel depicts protocol, green initialization scans generate an SiV⁻-rich background, a subsequent 3.6-mW, 3-s green park generates a SiV⁰ halo. For weak red readout powers, the dark halo is clearly visible, whereas at stronger powers (400 µW here) the SiV⁰ is rapidly converted to bright SiV⁻ during the scan and the dark halo vanishes. Each image is a composite, where a red readout image *without* a green park is subtracted from the image that follows the green park - the background of the 400 µW readout image appears significantly darker due to red bleaching of SiV and residual NV during the reference imaging step.

### V. Spurious Background Darkening/Brightening

Whenever high power (>1 mW) green light is parked for long durations (>1 s), we observe an unexpected phenomenon characterized by darkening or brightening of the image background depending on the initialization procedure. Similar effects have been reported in related experiments [5]. As an example, the web-like patterns of fluorescence observed in Fig. 4b of the main text around the SiV$^-$ ring and, more subtly, the brightening of the background in Fig 1b of the main text are seen consistently. We observed this artefact to varying degrees in several other samples with different NV/SiV concentrations, and eventually concluded it is optical in nature and unrelated to charge carrier generation and capture. Our leading hypothesis is that green light reflects from the bottom surface of the diamond during the laser park and undergoes several internal reflections. It has been shown that weak green illumination bleaches SiV$^-$ fluorescence and weakly re-pumps NV$^0$ into NV$^-$ [6], so the weak green retro-reflections inside the diamond lead to radially symmetric bleaching of SiV$^-$ or NV$^-$ generation over a wide area surrounding the laser park. Our conclusion that optical effects are to blame is based on two key observations:

1. The darkening or brightening effect does not appear to grow radially with park time or laser power. Instead, it appears to gradually reduce or increase the PL across typical fields of view of 100 µm or more as the green laser is held on, often with characteristically patchy, ring-like or tendril-like patterns suggestive of optical interference.
2. The effect is disrupted by degrading the back surface of the host diamond crystal, which stops reflections from the inner surface. For example, the rough surface finish that follows etching of patterns in the reverse surface with a laser cutter was found to locally eliminate the effect within the narrow-cut region.

| | | | |
|---|---|---|---|
| $P, Q, S$ | Ns, NV and SiV density | $1, 0.03, 0.3$ ppm | - |
| $\{P_+, Q_-, S_-, S_{2-}\}(r, 0)$ | Initial $N_s^+$, NV$^-$, SiV$^-$, SiV$^{2-}$ density | $0.469, 0.021, 0.03, 0.24$ ppm | - |
| $\sigma_{Nn}$ | N$^+$ electron capture cross section | $3.1 \cdot 10^{-6}$ µm$^2$ | [1,2] |
| $\sigma_{Np}, \sigma_{NVp}$ | N$^0$, NV$^-$ hole capture cross section | $1.4 \cdot 10^{-8}, 3.1 \cdot 10^{-6}$ µm$^2$ | [1,2] |
| $\sigma_{pS-}, \sigma_{pS2-}$ | SiV$^-$, SiV$^{2-}$ hole capture cross section | $9 \cdot 10^{-8}, 6 \cdot 10^{-5}$ µm$^2$ | * |
| $\sigma_{nS0}$ | SiV$^0$ electron capture cross section | $1.4 \cdot 10^{-8}$ µm$^2$ | * |
| $\vartheta_N$ | N$^0$ photoionization rate | 15 Hz | [1] |
| $\vartheta_0$ | NV$^0 \rightarrow$ NV$^-$ photorecombination rate | 0.0046 Hz | [1] |
| $\vartheta_-$ | NV$^- \rightarrow$ NV$^0$ photoionization rate | 0.01 Hz | [1] |
| $\vartheta_{S0}$ | SiV$^0 \rightarrow$ SiV$^-$ photorecombination rate | 0.01 Hz | * |
| $\vartheta_{S-}$ | SiV$^- \rightarrow$ SiV$^{2-}$ photorecombination rate | 0.001 Hz | * |
| $\mu_n$ | Electron mobility in diamond | $2.4 \cdot 10^{11}$ µm$^2$/(V·s) | [2,3] |
| $\mu_p$ | Hole mobility in diamond | $2.1 \cdot 10^{11}$ µm$^2$/(V·s) | [2, 3] |
| $D_n = \frac{\mu_n k_B T}{e}$ | Electron diffusion coefficient in diamond | $6.1 \cdot 10^9$ µm$^2$/s | - |
| $D_p = \frac{\mu_p k_B T}{e}$ | Hole diffusion coefficient in diamond | $5.3 \cdot 10^9$ µm$^2$/s | - |

**Table S2**. List of parameters used in the calculation of the SiV charge state distributions in Fig. 1 of the main text. $T$ = 300 K; $k_B = 1.38 \cdot 10^{-23} \frac{J}{K}$ is the Boltzmann constant, and $e = 1.6 \cdot 10^{-19}$ C is the elementary charge. Last column cites references for the values taken from the literature (* denotes parameters approximated in this work, as described in the text).

More pertinent to the current manuscript is the consequences of this spurious imaging artefact on the observed charge generation and capture. In a few limited cases where the darkening effect could be suppressed, we observed no difference in the generated SiV$^-$ concentric structures. We thus conclude that while spurious and detrimental to the imaging quality, the background darkening effect does not play a major role in the generation or detection of SiV charge states.

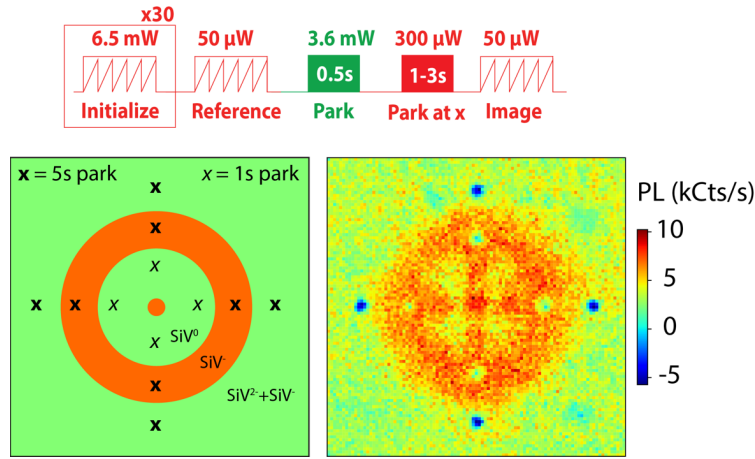

**Figure S7.** Protocol for initialization into a SiV$^{2-}$ rich charge state, green parking to generate an SiV$^0$/SiV$^-$ double halo and subsequent red parking to initiate photorecombination at 9 selected sites. Inside the SiV$^0$ region, we park a 300 μW beam for 1 s, whereas in the SiV$^-$ and SiV$^{2-}$ regions we park for 5 s. The bleaching observed in the SiV$^{2-}$ area is believed to be a result of residual SiV$^-$ not adequately eliminated by the initialization scanning.

**VI. SiV charge dynamics under red illumination**

At room temperature, SiV$^0$ ZPL fluorescence at 946 nm is far too weak to observe. As discussed in the main text, our inference of SiV$^0$ in the dark inner region after a green laser park is based on the direct observation of SiV$^0$ fluorescence at low temperatures (Fig 2 of the main text), the detection of carrier drift in the presence of an external electric field (Fig. 3 of the main text), and the characterization of the SiV fluorescence dynamics under weak red illumination at room temperature (Fig 4 of the main text). The key result of the latter method is the distinctly different behavior in the three different regions of SiV$^-$ fluorescence produced by the preceding green park. As shown in Fig. S7, parking in the inner region for short times selectively results in localized SiV$^-$ generation. Conversely, parking on the bright SiV$^-$ ring yields a dark charge state; we cannot regain SiV$^-$ fluorescence with subsequent weak red parks, which leads us to conclude the charge state is distinct in the inner region. Note that laser parking anywhere in the area surrounding the SiV$^-$ ring — assumed to be SiV$^{2-}$- rich — yields further bleaching. We interpret this observation with the aid of Fig. S3a, where repeated red scanning is found to deplete, though not totally, the SiV$^-$ charge state. The observed bleaching in the background region is thus considered to be a result of residual SiV$^-$ being locally extinguished. It is interesting nevertheless that the bleaching observed here is much more apparent compared to a red park on the SiV$^-$ bright ring. This may point to the role of other defects in the recombination process, as it appears the local charge environment affects the SiV$^-$ recombination rate.

At no point did we observe generation of a bright SiV$^-$ charge state in the area directly exposed to the red beam (whether weak or strong), other than within the dark inner ring region. We also examined NV$^-$-selective fluorescence under similar circumstances and concluded that all fluorescent activity reported in this work pertains only to SiV defects. Therefore, the distinctly different characteristics in the three different regions depicted in Fig. S7 can be correlated to the three stable charge states of the SiV defect.